\documentclass[preprint]{revtex4}
\usepackage{amsmath,amssymb} 
\usepackage{bm}
\usepackage{subfigure}
\usepackage{graphicx}

\begin{document}
\title{Field-sensitive addressing and control of field-insensitive neutral-atom qubits}

\author{N.~Lundblad} \email{nathan.lundblad@nist.gov}
\affiliation{Joint Quantum Institute, National Institute of Standards and Technology and University of Maryland, 
Gaithersburg, Maryland 20899, USA}

 \author{J.~M.~Obrecht}
\affiliation{Joint Quantum Institute, National Institute of Standards and Technology and University of Maryland, 
Gaithersburg, Maryland 20899, USA}

 \author{I.~B.~Spielman}
\affiliation{Joint Quantum Institute, National Institute of Standards and Technology and University of Maryland, 
Gaithersburg, Maryland 20899, USA}

 \author{J.~V.~Porto}
\affiliation{Joint Quantum Institute, National Institute of Standards and Technology and University of Maryland, 
Gaithersburg, Maryland 20899, USA}

\date{\today}
\begin{abstract}

The establishment of a scalable scheme for quantum computing with addressable and long-lived qubits would be a scientific watershed, harnessing the laws of quantum physics to solve classically intractable problems.   The design of many proposed quantum computational platforms is driven by competing needs: isolating the quantum system from the environment to prevent decoherence, and easily and accurately controlling the system with external fields.    For example, neutral-atom optical-lattice architectures provide environmental isolation through the use of states that are robust against fluctuating external fields, yet external fields are essential for qubit addressing.    Here we demonstrate the selection of individual qubits with external fields, despite the fact that the qubits are in field-insensitive superpositions.    We use a spatially inhomogeneous external field to map selected qubits to a {\em different} field-insensitive superposition (``optical MRI"), minimally perturbing unselected qubits, despite the fact that the addressing field is not spatially localized.      We show robust single-qubit rotations on neutral-atom qubits located at selected lattice sites.   This precise coherent control is an important step forward for lattice-based neutral-atom quantum computation, and is quite generally applicable to state transfer and qubit isolation in other architectures using field-insensitive qubits.
\end{abstract}

\maketitle

The ability to address individual qubits is a vital component of most quantum computing architectures.   In the case of neutral atom qubits held in an optical-lattice register~\cite{PhysRevLett.82.1060,PhysRevLett.82.1975,blochreview}, addressing generally requires the interaction of specific atoms with a control field.     Of long-standing concern is the difficulty of addressing only selected atoms amongst an ensemble of  $\approx 10^5$ atoms in nominally identical lattice sites.   One approach is to use long-period lattices or arrays of independent single-atom traps sufficiently spaced to address with a focused optical beam~\cite{Nelson:2007lr,Beugnon:2007fx,schrader:150501}; in this case the optical diffraction limit sets a bound on the qubit register spacing (and therefore register density).  Alternatively, experiments using the Mott insulator transition in optical lattices \cite{PhysRevLett.81.3108} result in large arrays of subwavelength-separated single ground-state atoms, which are useful for collisional and exchange quantum gates~\cite{PhysRevLett.82.1975,Mandel:2003tg} \cite{hayes:070501, Anderlini:2007qy}.       Schemes similar to magnetic resonance imaging for addressing subwavelength-separated qubits have been proposed and demonstrated, wherein an externally applied gradient field shifts local energies, mapping spectroscopic resolution to spatial resolution~\cite{schrader:150501,zhang:042316,lee:020402}.     However, these schemes require that the qubits be in field-sensitive states, a requirement which is at odds with the need for long coherence times.

\begin{figure}
\centering
  \includegraphics[]{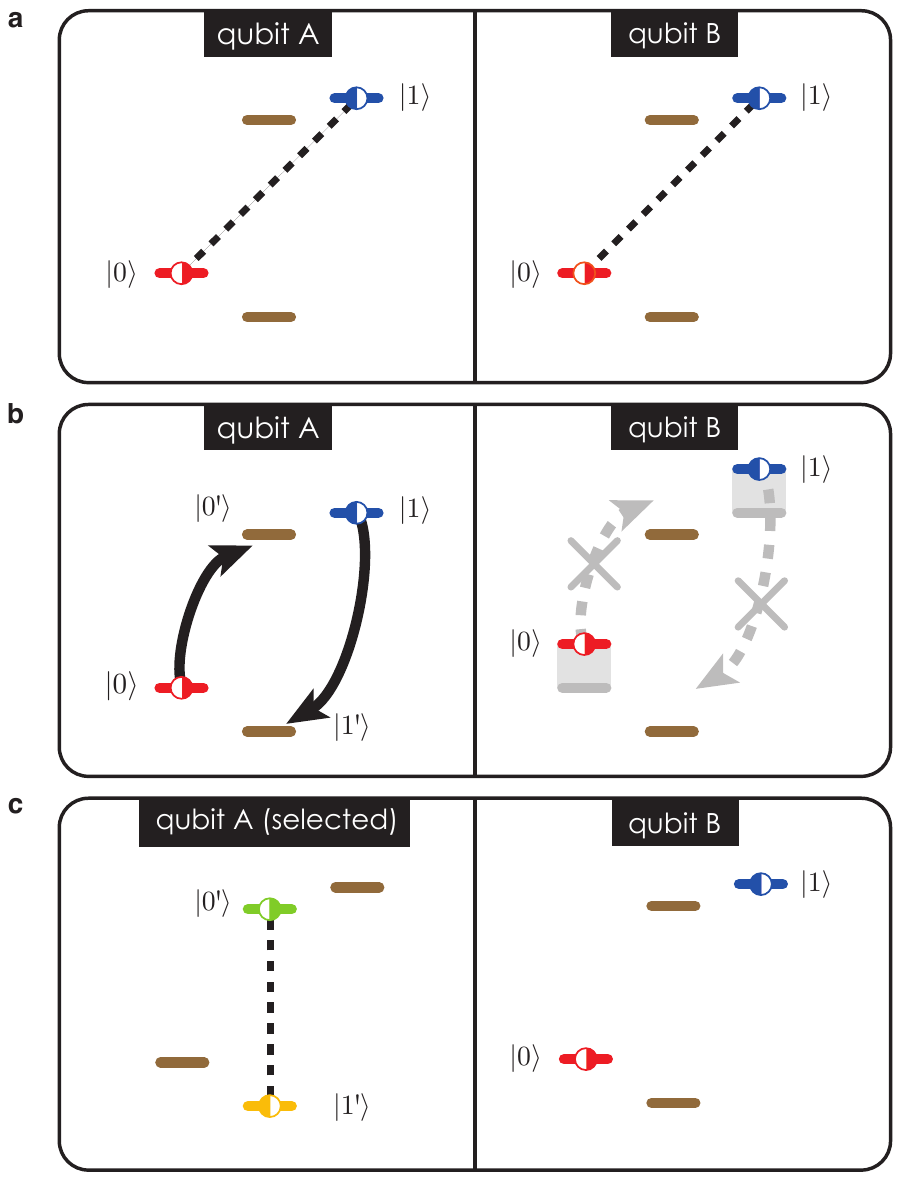}   
    \caption{{\bf Scheme for combining field-sensitive qubit addressability with long-lived field-insensitive ``clock state" qubits.}   {\bf a}, Using microwave or optical Raman control, qubits in a quantum register (two shown here as qubits A and B) can be prepared in a coherent superposition of the ``storage" states $|0\rangle$ and $|1\rangle$, a field-insensitive clock-state pair.    {\bf b},  Application of a magnetic field gradient Zeeman-shifts energy levels, spectrally selecting a specific qubit (A) from the register.      Field-sensitive transitions can then be used to selectively map the storage-state superposition to a pair of ``working" states $|0^\prime\rangle$ and $|1^\prime\rangle$.  {\bf c}, After the site-selective mapping, arbitrary qubit rotation can be performed on the A qubit alone, as the working-state transition is off-resonant from the storage-state transition.    Inverting the mapping process returns qubit A to a new storage-state superposition.   This scheme demonstrates crosstalk-free site-specific addressing, with both qubits almost always in field-insensitive superpositions.}    
\label{scheme}
\end{figure}

Here we demonstrate how to combine environmental insensitivity and site-specific subwavelength addressability, as illustrated in Fig.~\ref{scheme}.     Our scheme is based on a register of qubits, each in a superposition of {\it storage} states $|0\rangle$ and $|1\rangle$ whose energy difference is insensitive to external magnetic fields, a significant source of environmental decoherence.  (Such pairs of states are known colloquially as ``clock states" due to their utility as frequency standards.)   In addition, each qubit has a second pair of clock states, the {\it working} states $|0^\prime\rangle$ and $|1^\prime\rangle$, which have a different transition frequency than the storage states.     While transitions between the storage states ($|0\rangle\leftrightarrow |1\rangle$) or between the working states  ($|0^\prime\rangle\leftrightarrow |1^\prime\rangle$) are insensitive to external fields, the transitions between storage and working states (e.g. $|0\rangle\leftrightarrow |0^\prime\rangle$) {\em are} field-sensitive.     Application of a field gradient thus spectrally selects a qubit from the register, allowing frequency-sensitive (and therefore position-sensitive) mapping of the selected qubit's coherent superposition between storage states and working states.   Once the selected qubit has been transferred to the working states, we can then perform isolated arbitrary single-qubit operations on the selected qubit alone. Remaining qubits (still in the storage states) are unaffected by both the mapping process and the subsequent control operation on the selected qubit.    One can then map the selected qubit back to storage states, resulting in a qubit register with the addressed qubit in a new arbitrary superposition of storage states.   This idea may be applicable to other physical systems vulnerable to optical or electrical  crosstalk (e.g., Refs.~\cite{PhysRevLett.83.4204} and \cite{Taylor:2005zm}), relaxing the required isolation for control fields used to address individual qubits in a spatially dense register.

   We demonstrate this scheme in an optical lattice-based ensemble of  registers, where each register is composed of two separately trapped $^{87}$Rb atoms acting as qubits A and B.   The storage and working states of the qubit are encoded in four hyperfine sublevels of the ground-state manifold of $^{87}$Rb, which can be coupled with resonant microwave radiation.   Both qubits are initialized in the storage-state superposition  $\alpha|0\rangle+\beta|1\rangle$.   After applying a localized effective magnetic field, we spectrally select qubit A and map it into the working-state superposition  $\alpha |0^\prime\rangle+\beta e^{i \theta} |1^\prime\rangle$, where $\theta$ is a systematic phase depending on the details of the mapping process.      Qubit B remains in the initial storage-state superposition.  We then apply a control operation on the working transition, transforming qubit A into the new state   $\alpha^\prime  |0^\prime\rangle+\beta^\prime |1^\prime\rangle$.    Using a modified Ramsey technique we verify the basic features of this scheme: namely, that storage-state coherence is unaffected by the application of the field gradient, and that the addressable mapping process is coherent.

\vspace{.5in}

\begin{figure}
\centering
  \includegraphics[width=\columnwidth]{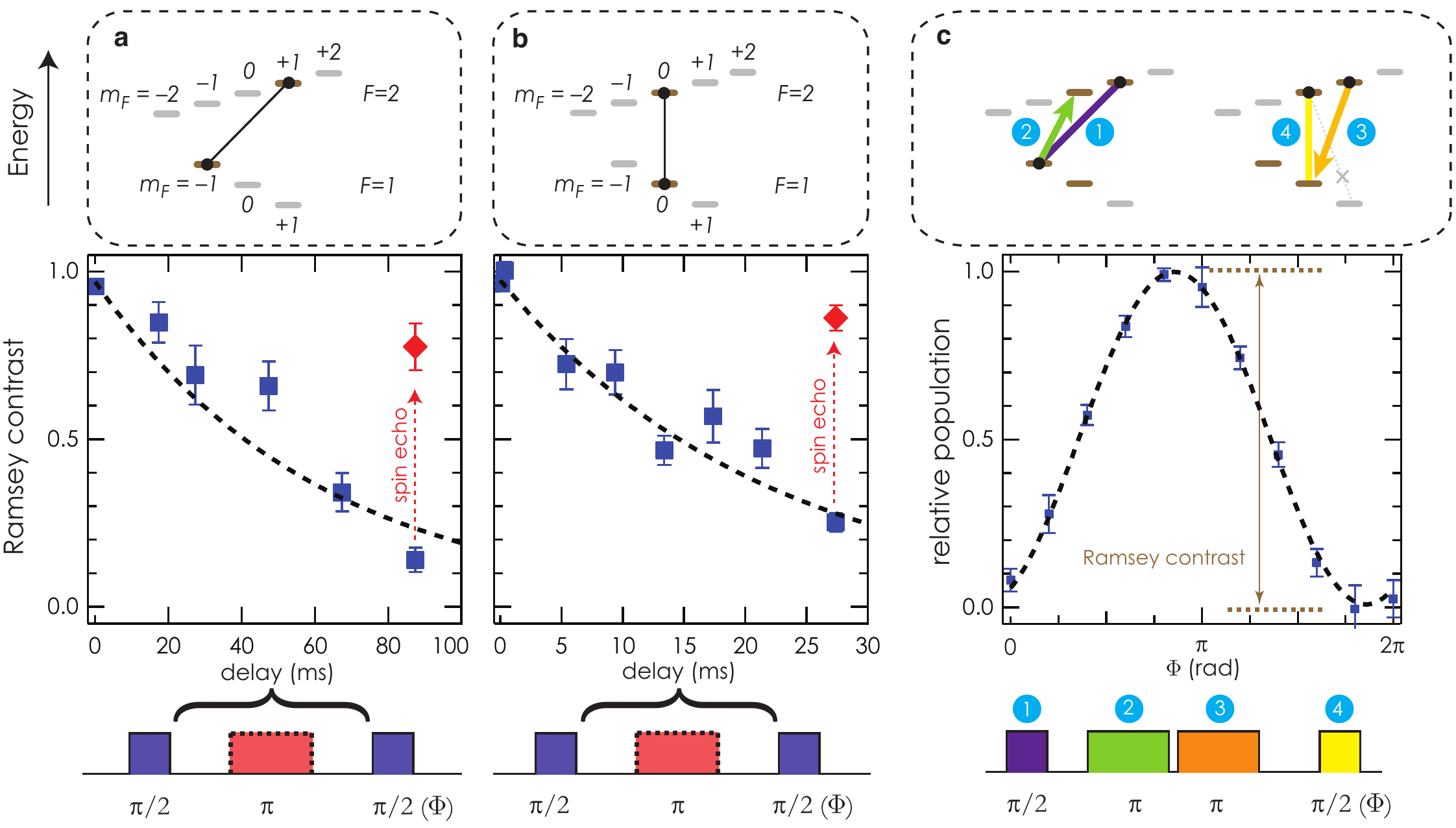}   
    \caption{
    {\bf Qubit coherence and qubit mapping.} {\bf a}, We measure the coherence time  $T_2^*$ of the storage-state qubit.   The contrast of the resulting interference fringe as a function of delay is shown; the storage-state qubit exhibits a coherence time of $\simeq$ 61(5) ms, assuming exponential decay.  The insertion of a spin-echo $\pi$-pulse in the Ramsey sequence removes dephasing caused by spatial inhomogeneities, resulting in longer coherence times (red).   {\bf b},  A similar measurement, performed using the working states, yields a coherence time of $\simeq$ 21(2) ms, which also improves given a spin-echo $\pi$-pulse (red). {\bf c}, We demonstrate the coherent mapping of a storage-state superposition to a working-state superposition.   A modified Ramsey pulse sequence opens on the storage transition (purple) and closes on the working transition (yellow), giving  a fringe of observed contrast of 0.99(3).   The mapping is effected with the use of appropriate $\pi$-pulses transferring the $|0\rangle$ population to $|0^\prime\rangle$ (green) and the $|1\rangle$ population to $|1^\prime\rangle$ (orange).   We control undesired ``leakage" to a nearby state, as illustrated with the dashed grey transition.}    
\label{coherence}
\end{figure}

The confinement for our register is provided by a double-well optical lattice (see Methods and Ref.~\cite{sebby-strabley:033605}).   We use the states $|F=1,m_F=-1\rangle$ and $|F=2,m_F=1\rangle$ as storage states $|0\rangle$ and $|1\rangle$, and the states  $|F=1,m_F=0\rangle$ and $|F=2,m_F=0\rangle$ as working states $|1^\prime\rangle$ and $|0^\prime\rangle$.    At our operating field near 323 $\mu$T, the linear magnetic-field dependence of the storage transition vanishes, making the storage-state qubit extremely insensitive to small field fluctuations or inhomogeneity~\cite{PhysRevLett.81.243}.  The working transition has a slight linear sensitivity to magnetic field.       Figure~\ref{coherence}(a-b) shows the coherence properties of the storage and working states, measured using Ramsey's method of separated oscillatory fields~\cite{PhysRev.78.695}.     Fig.~\ref{coherence}(a) shows the result of a standard Ramsey sequence, where we ``open"  with an initial $\pi/2$-pulse on the storage transition, placing each qubit in the equal superposition $|0\rangle + |1\rangle$.   After a fixed delay, we ``close" the sequence with a final $\pi/2$-pulse of variable relative phase $\Phi$, and measure the populations in $|0\rangle$ and $|1\rangle$ as a function of $\Phi$.   The resulting fringe contrast as a function of the delay time decays with a dephasing time $T_2^*=61(5)$ ms.

       The residual dephasing is limited by inhomogeneity of the lattice-beam intensity since there is no known simple ``magic wavelength" scheme ~\cite{PhysRevLett.91.173005} for hyperfine transitions in $^{87}$Rb~\cite{rosenbusch:013404,flambaum:220801}.  This inhomogeneity results in a small, lattice-induced differential light shift of the hyperfine states that is proportional to lattice intensity and inversely proportional to the lattice-beam detuning from atomic resonance.    We calculate a differential  shift of $\simeq 170$ Hz  for a typical total lattice light shift of 230 kHz  and estimate an inhomogeneity of $\simeq$ 10 Hz  for our experimental parameters, which is roughly consistent with the observed dephasing time.   We also measure the magnitude of the differential shift as a function of lattice intensity (discussed below).  The inhomogeneity of the  differential light shift is the main technical limitation on coherence in our ensemble of storage qubits, and can be improved by making the lattice beams more homogeneous and detuning the lattice farther from resonance, as with the $\lambda = 1.06$ $\mu$m lattice used for precision spectroscopy in Ref.~\cite{Campbell08042006}.     In addition, dynamical decoupling techniques have been shown to be useful in reducing the effects of inhomogeneous broadening~\cite{biercuk2008odd,uhrig:100504}.   A simple spin-echo pulse sequence, which rephases time-independent inhomogeneous dephasing, gives a residual coherence time $T_2$ in excess of 300~ms.   In an analogous measurement, Fig.~\ref{coherence}(b) shows that the working-state coherence exhibits a somewhat shorter dephasing time of $T_2^*= 21(2)$ ms, dominated by background magnetic field gradients.

In contrast to the field insensitive qubit states, the transitions $|0\rangle \rightarrow |0^\prime\rangle$ or $|1\rangle\rightarrow |1^\prime\rangle$ {\em are} field sensitive.  This sensitivity  enables spectroscopic qubit addressing, but  requires that state transfers be performed fast enough to avoid the qubits' vulnerability to field inhomogeneities while they temporarily occupy the field-sensitive superpositions $|0^\prime\rangle+|1\rangle$ or $|0\rangle+|1^\prime\rangle$ during the transfer from storage states to working states, as in Fig.~\ref{scheme}(b).    We measure this timescale using the standard Ramsey method on the field-sensitive transition $|0\rangle \rightarrow |0^\prime\rangle$, giving  $T_2^*  \simeq 500$  $\mu$s,  roughly 100 times shorter than that of the clock states.     While the Ramsey dephasing is linearly sensitive to weak field inhomogeneity, the transfer efficiencies of the $\pi$-pulses in our mapping sequence are only quadratically sensitive.   This suggests that the mapping is less sensitive to field inhomogeneities than simply given by $T_2^*$.    To directly determine the impact of this field sensitivity on the mapping from storage to working states, we developed a modified Ramsey sequence that is sensitive to decoherence in the mapping process.     This sequence is composed of the standard opening $\pi/2$-pulse on the storage transition, followed by $\pi$-pulses mapping storage states to working states, and closed by a final $\pi/2$-pulse on the working transition.      While the standard Ramsey sequence depends only on the relative phase of the two (equal-frequency) $\pi/2$-pulses, our modified sequence requires phase control of four different microwave signals, including the two mapping $\pi$-pulses.     Because all our pulses have different frequencies, the meaning and control of the relative phases involved is more subtle.  

In particular, each modified Ramsey sequence (performed at time $t_0$ relative to a fixed origin) involves four signals of the form $\cos{(\omega t+ \phi)}$, and  the observed fringe depends on the relative phase 
$\phi_{\rm tot} = \phi_{01}+\phi_{0^\prime 1^\prime}-\phi_{00^\prime}-\phi_{11^\prime} + \Delta\omega t_0$, where 
\begin{equation}\label{sum}
\Delta\omega=  \omega_{01}+\omega_{0^\prime 1^\prime}-\omega_{00^\prime}-\omega_{11^\prime}.
\end{equation}
 If the four frequencies are phase-locked and satisfy the energy-conserving condition $\Delta\omega=0$,  the Ramsey output is insensitive to the starting time $t_0$ of the sequence, depending only on $\phi_{\rm tot} = \phi_{01}+\phi_{0^\prime 1^\prime}-\phi_{00^\prime}-\phi_{11^\prime}$.    We observe Ramsey fringes  by adjusting the phase of any of the microwave fields by a variable offset $\delta\phi$, with the $\phi_{01}$ and $\phi_{0^\prime 1^\prime}$ fringes out of phase with the  $\phi_{00^\prime}$ and $\phi_{11^\prime}$  fringes.     A fringe resulting from varying $\phi_{0^\prime 1^\prime}$ is shown in Fig.~2(c).   Despite the fact that the total duration of the mapping process (200 $\mu$s) is comparable to the dephasing time of the sensitive transitions, the fringe has high contrast, confirming that the qubit coherence is largely unaffected by the mapping process.   

\begin{figure}
\centering 
  \includegraphics[width=\columnwidth]{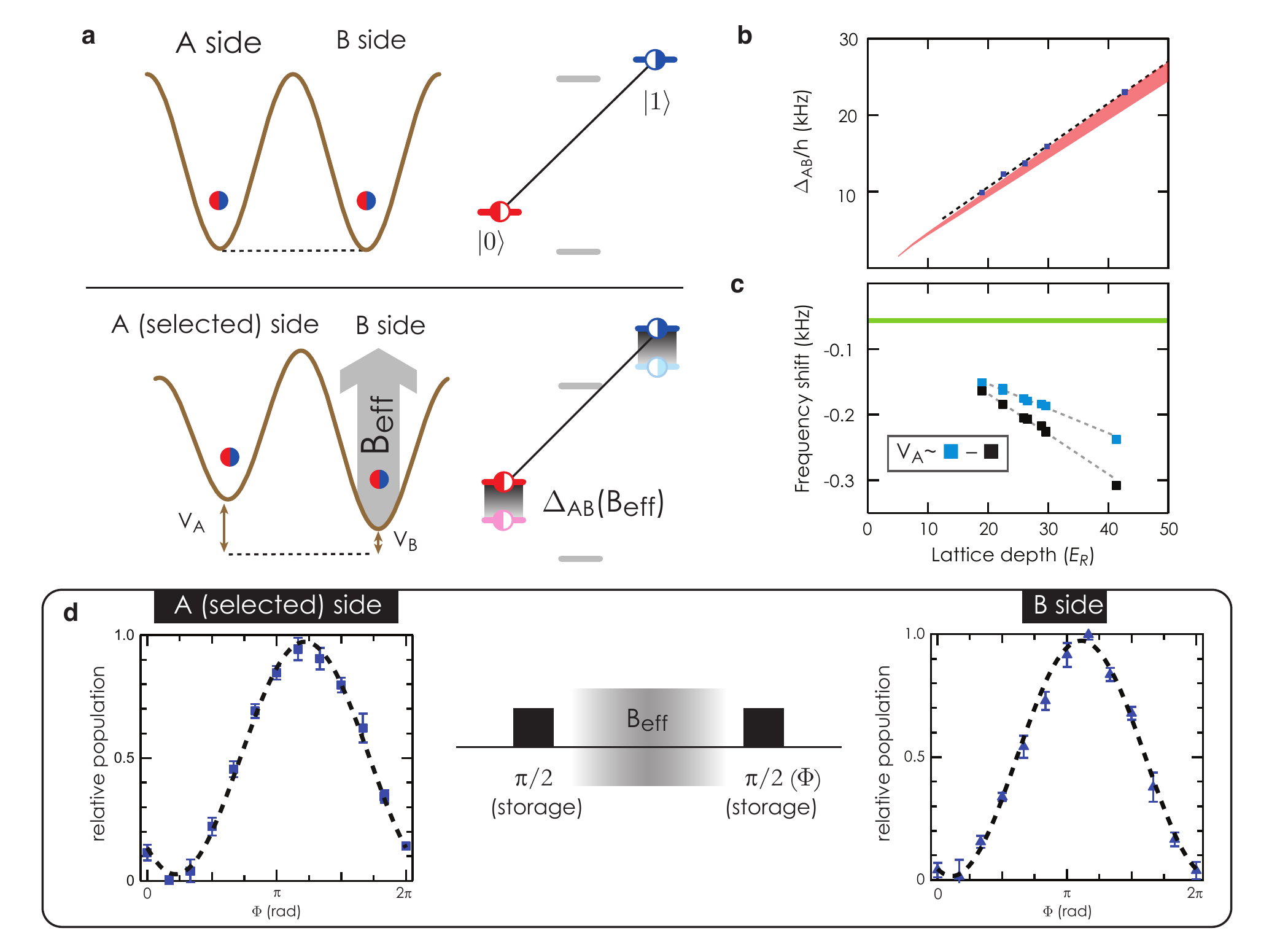}
      \caption{   {\bf $\vec{\bf B}_{\bf eff}$ and the differential shift.}   {\bf a},  We apply an effective magnetic field $\vec{B}_{\rm eff}$ to the B sites of a two-qubit register, Zeeman-shifting the field-sensitive levels by an amount $\Delta_{\rm AB}$.     Additionally, in the $\vec{B}_{\rm eff}\neq 0$ configuration the light shift at each site changes slightly by amounts $V_A$ and $V_B$.   {\bf b},   The calibration of $\Delta_{\rm AB}$ is shown as a function of lattice-beam intensity, where  the intensity is plotted in units of the equivalent $\vec{B}_{\rm eff}=0$ lattice depth (see Methods).     A prediction of the frequency shift based on a model of our lattice is shown in pink.   {\bf  c}, Measurements of the differential shift of the working-transition resonance frequencies in the $\vec{B}_{\rm eff}=0$ lattice (black) and  on the A sites of the $\vec{B}_{\rm eff}\neq 0$ lattice (blue) are shown as a function of intensity, relative to the expected free-space resonance, effectively measuring $V_A$.    Shown in green is an estimate of the differential shift caused by the lattice along $\hat{z}$, which is held at constant intensity.       {\bf d}, Ramsey fringes are shown illustrating the  coherence of storage-state qubits after application of $\vec{B}_{\rm eff}$ to the B sites during part of the Ramsey delay.   We observe Ramsey fringes for the A sites (left graph, contrast 0.95(2)) and B sites (right graph, contrast 0.96(2)).}
\label{leftright}
\end{figure}

We obtain the necessary field gradients to make the mapping process addressable by generating an optically-induced effective magnetic field at every other lattice site, i.e., on one site of each of our two-qubit registers, as illustrated in Fig.~\ref{leftright}(a).    This optically-induced effective magnetic field is proportional to the local ellipticity of the lattice light, which we control using electro-optic modulators (see Methods and Ref.~\cite{sebby-strabley:033605}).   The spatially varying effective magnetic field $\vec{B}_{\rm eff}$ results from the atom's vector light shift (which adds to the scalar light shift providing the lattice potential), where $\vec{B}_{\rm eff}\sim i \alpha_v (\vec{E}^*\times\vec{E})$, $\alpha_v$ is the vector polarizability, $\vec{E}$ is the lattice electric field, and $\vec{E}^*\times\vec{E}$ is the local ellipticity of the lattice field.     $\vec{B}_{\rm eff}$ adds vectorially with the external bias field~\cite{deutsch2000qtm,sebby-strabley:033605,lundblad:150401,PhysRevLett.91.010407}.  We can adiabatically transform the initial $\vec{B}_{\rm eff}=0$ lattice  into a lattice with a nonzero $\vec{B}_{\rm eff}$ at the B sites of each quantum register, Zeeman-shifting the resonant frequency of the B sites by  $\Delta_{\rm AB}/h$.    The effective field remains zero at the A sites, resulting in an effective magnetic field gradient between A and B sites.   For a given polarization configuration, this shift is proportional to the lattice-beam intensity as illustrated in Fig.~\ref{leftright}(b), and corresponds to a field gradient of up to $\simeq$ 8 T/m.      In the presence of this gradient, the A  and B sites of the quantum register can be addressed using radiofrequency or microwave fields.   We now demonstrate a) that our qubits are largely insensitive to $\vec{B}_{\rm eff}$, yet b) we can nevertheless use $\vec{B}_{\rm eff}$ to coherently address selected qubits in our quantum register. 
    
The robustness of the storage-state qubit in the presence of the effective magnetic field is illustrated in Fig.~\ref{leftright}(d).   We show high-contrast Ramsey fringes, measured independently for A and B sites, where we apply the addressing field $\vec{B}_{\rm eff}$ to the B qubits between the $\pi/2$-pulses comprising the Ramsey sequence.     As expected, the storage-state coherence on either site is unaffected by $\vec{B}_{\rm eff}$.     During the application of $\vec{B}_{\rm eff}$, a phase offset of 19(2)$^\circ$ develops between the Ramsey fringes of the two sites, which corresponds to a small energy difference of $h\times$35 Hz over the 1.5 ms Ramsey delay.     This partially results from an intensity difference (and associated difference in differential light shifts) that exists between the A and B sites in the $\vec{B}_{\rm eff}\neq 0$ configuration (see Fig.~\ref{leftright}(a)).     In order to understand this, we measure the magnitude of the differential light shift for the working-state transition, first in the $\vec{B}_{\rm eff}=0$ configuration, and then for the A sites in the  $\vec{B}_{\rm eff}\neq 0$ configuration, both as a function of intensity, as illustrated in Fig.~\ref{leftright}(c).       We parametrize the lattice-beam intensity in all lattice configurations in terms of the equivalent lattice depth of the $\vec{B}_{\rm eff}=0$ configuration (see Methods).   Given this parametrization,  we measure the differential shift for atoms in the $\vec{B}_{\rm eff}=0$ configuration to be 6.1(5) Hz/$E_R$, where $E_R= \hbar^2 k^2/2 M = h\times 3.499$ kHz, with $k = 2 \pi/\lambda$ and M is the $^{87}$Rb atomic mass.   This is near a calculated value of 4.9 Hz/$E_R$ based on a model of the lattice and an atomic light-shift calculation.   The difference between these curves represents crosstalk: the extent to which the addressing process perturbs A-site atoms, which nominally experience no effective magnetic field.    A typical 25 Hz difference in differential light shifts on the A sites (between the $\vec{B}_{\rm eff}= 0$ and $\vec{B}_{\rm eff}\neq 0$ configurations ) combined with a typical 15 kHz effective Zeeman shift of the B sites suggests a crosstalk figure-of-merit for our system of $\simeq$~0.002.      For the two storage-state Ramsey  fringes of Fig.~\ref{leftright}(d), the estimated difference in differential shifts between the A and B sites of $\simeq$ 12 Hz (different than that of the case of the working transition) combined with the $\vec{B}_{\rm eff}$ application time of $\simeq$ 600 $\mu$s gives an expected phase shift of $\simeq$ 3$^\circ$, considerably smaller than that observed.      Several effects could contribute to this discrepancy: drifts in the lattice intensity, peculiarities of beam alignment not included in our lattice model~\cite{sebby-strabley:033605}, and drifts in both microwave/rf power and background magnetic fields leading to shifts in the two-photon transition controlling the storage-state qubits (see Methods).

\begin{figure}
\centering
  \includegraphics[]{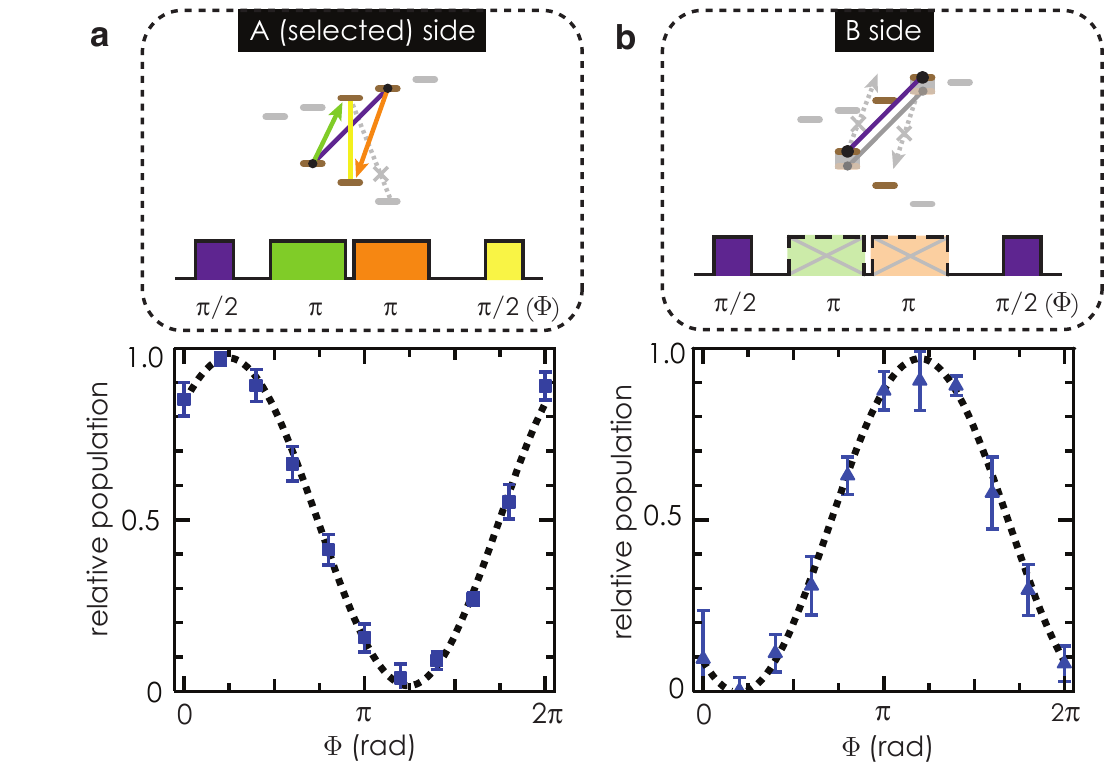}   
    \caption{{\bf Site-selective coherent mapping and single-qubit rotation.}     We open a Ramsey sequence  on the storage transition on both A and B sites, perform the storage-state to working-state mapping on A sites alone, and then {\bf a}, close the Ramsey sequence on the working transition, and measure state population on A sites alone with observed fringe contrast 0.96(2.5); {\bf b}, close the Ramsey sequence on the storage transition, measuring state population on B sites alone with observed fringe contrast 0.97(4).     A--B isolation and leakage to undesired states are controlled via tailored Rabi frequencies on the mapping pulses, and the appropriate choice of $\vec{B}_{\rm eff}$(see text). }    
\label{precision}
\end{figure}

Figure~\ref{precision} illustrates our full capability, where we combine coherent mapping between storage and working states with the addressing provided by the use of $\vec{B}_{\rm eff}$.    The combination is complicated by the need to simultaneously satisfy three possibly conflicting criteria: the prevention of  ``leakage" of the mapped-site population into unwanted hyperfine states (see Fig.~\ref{precision}(a)), and the assurance that each of the two mapping pulses affects only one site (A--B isolation: see Fig.~\ref{precision}(b)).    As our Rabi frequencies are comparable to both the effective Zeeman shift from $\vec{B}_{\rm eff}$ as well as shifts of the hyperfine states due to the nuclear magnetic moment, some care is required.     We initialize both qubits in the register with a $\pi/2$-pulse on the storage-state transition, apply the effective magnetic field to the B sites, then apply the mapping pulses.    Since  $\vec{B}_{\rm eff}$  shifts the B-site mapping transitions from the A-site resonance (by $\Delta_{\rm AB}/h=23$ kHz),  conversion from storage states to working states is performed {\em only} on the A sites.     We satisfy the above three isolation criteria by appropriately choosing three mapping parameters: the effective Zeeman shift $\Delta_{\rm AB}$ and the two microwave Rabi frequencies of the mapping $\pi$-pulses.  In particular, the Rabi frequency of the second mapping pulse ($|1\rangle\rightarrow|1^\prime\rangle$) was chosen to eliminate leakage to undesired states, as in the global operation of Fig.~\ref{coherence}; additionally, we chose $\Delta_{\rm AB}$ such that the second mapping pulse exhibited zero response at a detuning of $\Delta_{\rm AB}/h$.   Finally, we chose the Rabi frequency of the first mapping pulse ($|0\rangle\rightarrow|0^\prime\rangle$) so that it similarly exhibited zero response at a detuning of $\Delta_{\rm AB}/h$ .   

To verify the site-selective mapping of the A-site atoms to the working states, we a) measure the state population in the A sites after closing the Ramsey sequence with a $\pi/2$-pulse on the working transition, and b) measure the state population in the B sites after a $\pi/2$ pulse on the storage transition.     The data shown in Fig.~\ref{precision} proves the coherence and isolation of the transfer process and also demonstrates controlled single-qubit rotation performed on only one site of the register.    The high contrast of the resulting fringes confirms that the process is coherent, and that state pollution from imperfect A--B isolation or intra-site leakage to undesired states is at most  3\% (comparable to our measurement uncertainty of about 3\%).       

\begin{figure}[t]
\centering
  \includegraphics[]{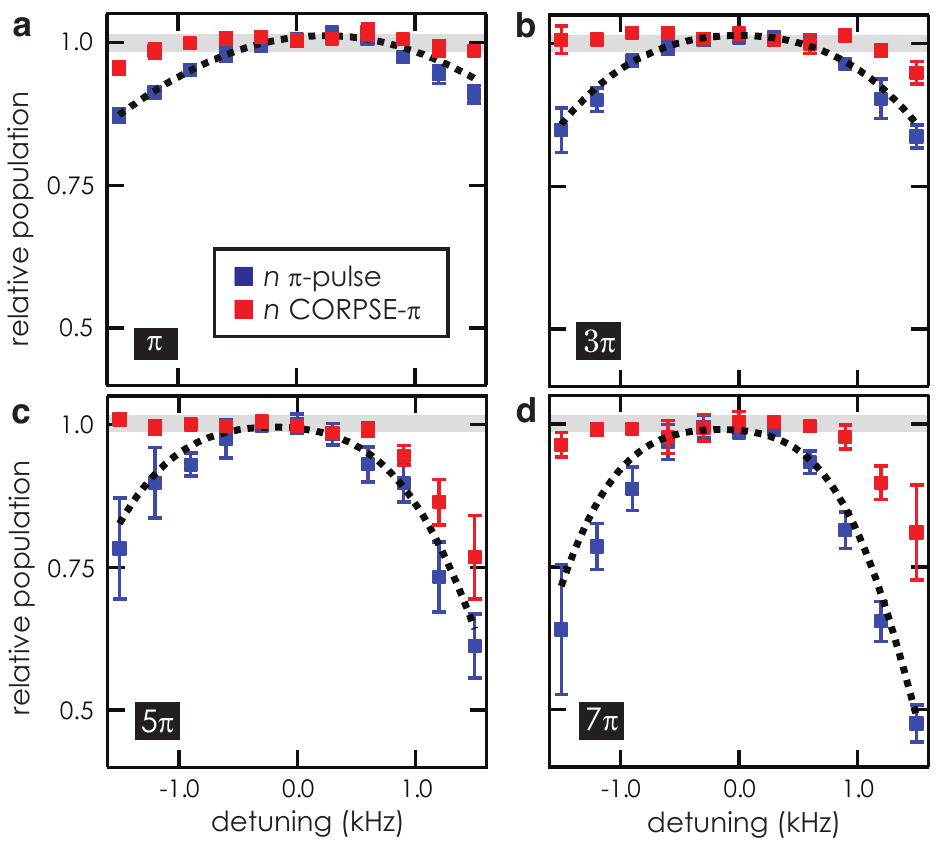}   
    \caption{{\bf Exploration of composite-pulse techniques for trapped neutral atoms.}  We compare the frequency spectrum  of $n$ $\pi$-pulses vs $n$ CORPSE-$\pi$ pulses on a field-sensitive transition.  The $\pi$-pulses (denoted as $180_x$) have arbitrary phase and standard pulse area, while the CORPSE pulses comprise three sequential pulses of differing area and phase, in this case $420_0 300_{180} 60_0$.      The $n\pi$ analogue of a CORPSE pulse is simple $n$-fold repetition of the CORPSE pulse.      {\bf a-d}, data representing $\pi$, $3\pi$, $5\pi$, $7\pi$-pulses and their respective CORPSE analogues.   The center frequency of the resonance drifts slightly between between graphs, illustrating the errors to which these experiments are vulnerable.  The dashed lines represent the expected spectrum of an $n\pi$-pulse given the measured Rabi frequency of the transition. }\label{composite}
\end{figure}

Any control scheme incorporating field-sensitive transitions (as ours does) is vulnerable to imperfections in the quality of the control, due to fluctuating or inhomogeneous background or control fields.    While our observed global Ramsey fringe contrasts are consistent with unity, scalable quantum computing places stringent limits on required control fidelities.    This level of control can be achieved with composite pulses, providing a specific desired result (such as that of the $\pi$-pulses used in our transfer process) using a train of pulses of variable pulse area and phase.   Composite pulses are designed to be robust against fluctuations and inhomogeneities over a given bandwidth~\cite{levitt1986cp}; pulses of particular interest to the quantum computing community have been discussed~\cite{PhysRevA.67.042308} and explored experimentally~\cite{rakreungdet2008amc}.      As a proof of principle, Fig.~\ref{composite} shows the results of applying the venerable CORPSE pulse sequence on the field-sensitive transition  $|F=1,m_F=-1\rangle\rightarrow|F=2,m_F=0\rangle$, along with the results of conventional $\pi$-pulses.   As expected, the $n$-CORPSE-$\pi$ spectra are significantly flatter about resonance than the equivalent $\pi$-pulse spectra.  These are not immediately applicable to our transfer process due to issues involving A--B isolation, but demonstrate the inherent utility of the technique for neutral atom quantum computation.    

The technique demonstrated here can be used with the effective field gradient of an individual focused laser beam \cite{zhang:042316} to provide single site addressing. The ultimate fidelity of such addressable control will be determined by a range of technical issues, such as the stability of the external and control fields, the spatial resolution of the addressing beam and its registration to the lattice position, and the lattice intensity. The fundamental limit is set by the spontaneous photon scattering of the light providing the effective magnetic field gradient. In $^{87}$Rb atoms, for example, one optimal choice for the addressing laser wavelength is found near 787 nm, detuned between the $6^2 P_{1/2}$ and $6^2 P_{3/2}$ transitions. The total scattering probability during a site-selective $\pi$-pulse depends on the details of the system \cite{zhang:042316} but we calculate that for a 1~$\mu$m beam waist and 0.5~$\mu$m lattice spacing, the scattering probability can be $\lesssim 2\times10^{-4}$. In the context of our double-well lattice, future work will focus on implementing the transfer process with composite pulses and implementing benchmarking techniques~\cite{knill:012307} to probe our control fidelities below the percent level. We also seek to use our techniques to provide all possible inputs to the $\sqrt{ \rm SWAP}$ gate described in Ref.~\cite{Anderlini:2007qy}, where the relevant gate time is more than two orders of magnitude faster than the relevant coherence time. In a more general context,  the approach we have demonstrated here is applicable to any quantum computing  architecture where two appropriate pairs of states can be found, with the attendant ability to transform in and out of a storage-state quantum memory.

\section{Methods}

  We load our optical lattice with ultracold atoms originating from a spin-polarized $^{87}{\rm Rb}$ Bose-Einstein condensate (BEC) in the $5^2 S_{1/2}$ $|F=1,m_F=-1\rangle$ hyperfine ground state.     We produce small condensates with $\approx 8\times 10^4$ atoms (such that the resulting lattice filling factor is near unity) in a Ioffe-Pritchard magnetic trap; to this trap we subsequently add a three-dimensional optical lattice, generated by intersecting beams from a  Ti:sapphire laser operating at $\lambda=810.0$ nm.   A lattice along $\hat{z}$ divides the BEC into a stack of independent 2D systems, and a separate, deformable double-well lattice  formed from a single folded and retroreflected beam in the $xy$ plane completes the confinement\cite{sebby-strabley:033605}.     Since the total lattice depth in the lattice is a complicated function of the laser polarizations controlling the topology, we parametrize the intensity of the single $xy$-lattice input beam for all lattice configurations in terms of the equivalent $xy$-lattice depth in the $\vec{B}_{\rm eff}=0$ configuration.     In terms of this parametrization, the total light shift experienced by a trapped atom in the $\vec{B}_{\rm eff}=0$ configuration will be approximately twice the $xy$-lattice depth plus the depth of the lattice along $\hat{z}$.   During loading, all lattice intensities follow an exponential profile, reaching their final values in 150 ms (with a time constant of 50 ms).    Typical final lattice depths are 20(1) $E_R$ for the vertical lattice and 20(1) to 40(2) $E_R$ for the $xy$-lattice, where $E_R= \hbar^2 k^2/2 M = h\times 3.499$ kHz, with $k = 2 \pi/\lambda$ and M is the $^{87}$Rb atomic mass.    (Unless otherwise stated, all uncertainties herein reflect the  uncorrelated combination of  statistical and systematic uncertainties.)  For all experiments, the initial lattice we load into is a state-independent square lattice of period $\lambda/2= 405 $ nm.    In Fig.~3(c), the points at 41 $E_R$ were obtained at a lower depth of the lattice along $\hat{z}$, and have been corrected by 18 Hz (the estimated change in the differential shift from the lattice along $\hat{z}$) to be consistent with the other data.
  
Atoms loaded into this initialization lattice are deep into the Mott-insulating phase with nominally one atom per site.  We then turn off the Ioffe-Pritchard trap, leaving a stable bias field of 322.9(1) $\mu$T along $\hat{x}-\hat{y}$.     This field is chosen to minimize the sensitivity of the hyperfine clock transition $|1,-1\rangle\leftrightarrow |2,+1\rangle$ (our storage qubit) to external fields, while allowing for reasonably low sensitivity ($\simeq$ 37 kHz/mT) in our working qubit, the well-known $|1,0\rangle\leftrightarrow|2,0\rangle$ transition commonly used in atomic clocks.     

We transform our lattice into the state-dependent configuration ($\vec{B}_{\rm eff}\neq 0$) on a timescale (300 $\mu$s) adiabatic with respect to vibrational excitation.  This transformation is effected through the use of high-voltage Pockels cells which control both the input polarization of the $xy$-lattice, and the relative phase of the in-plane ($\hat{x}$, $\hat{y}$) and out-of-plane ($\hat{z}$) polarization components~\cite{sebby-strabley:033605}.  This yields an A/B site dependent ellipticity which in concert with the atoms' vector polarizability yields a spin-dependent vector light shift.    To measure the state population of  the A(B) lattice sites, we dynamically adjust the topology of the lattice, converting B(A)-site atoms into high momentum states which spatially separate from the low-momentum atoms of interest in the A(B) sites during time-of-flight~\cite{lee:020402}.   

We measure the atomic density distribution using resonant absorption imaging along $\hat{z}$ and use Stern-Gerlach gradients~\cite{Gerlach1922} to concurrently resolve differing $m_F$ components.     In addition to standard technical noise, uncertainty in the contrast measurements in the A(B) site is due in part to imaging noise from background atoms from the B(A) site not being intentionally measured.   Depending on the depth of the lattice, the high-momentum atoms may spatially overlap with low-momentum atoms in the images;  the data in Fig.~\ref{leftright} (for which this effect was slight) a small correction was applied.      The lattice depth used in Fig.~\ref{precision} was large enough (chosen for maximum A--B isolation) that the high-momentum states did not overlap, and no correction was necessary.

Transitions between storage states are driven by a two-photon coupling comprising a microwave field at 6832.325 MHz and a rf field at 2.352975 MHz~\cite{PhysRevLett.81.243}.   For our field intensities and detunings  the two-photon Rabi frequency is $\Omega_{2\gamma}/2\pi\simeq 750(8) $ Hz, given a $\simeq 100$ kHz detuning from the $|F=2,m_F=0\rangle$ intermediate state.    Unlike single-photon transitions, the two-photon transition is sensitive to power-dependent shifts due to the radiofrequency and microwave coupling to the intermediate state.   This shift is only present during the application of the coupling, and must be considered when satisfying the the energy conserving condition of Eq.~\ref{sum}.    This shift is typically $\simeq$ +50 Hz, however, and the $\omega_i$ can be adjusted to satisfy Eq.~\ref{sum} without compromising resonance, as the single-photon Rabi frequencies $\Omega_i/2\pi$ are all $>$ 4 kHz.    

\section{Acknowledgements}
We thank Poul Jessen and Ivan Deutsch for helpful discussion, and Steven Swift and Eric Huang  for technical assistance with our direct-digital-synthesis hardware.  This work was partially supported by DTO and  ONR.  N.L. acknowledges support from the National Research Council Research Associateship program. 

\bibliographystyle{apsrevNOURL}

\pagebreak

%

\end{document}